# DO ELECTROMAGNETIC WAVES EXIST IN A SHORT CABLE AT LOW FREQUENCIES? WHAT DOES PHYSICS SAY?

HSIEN-PU CHEN [1], LASZLO B. KISH [1], CLAES-GÖRAN GRANQVIST [2], GABOR SCHMERA [3]

[1] *Texas A&M University, Department of Electrical and Computer Engineering, College Station, TX 77843-3128, USA*

[2] *Department of Engineering Sciences, The Ångström Laboratory, Uppsala University, P.O. Box 534, SE-75121 Uppsala, Sweden*

[3] *Space and Naval Warfare Systems Center, San Diego, CA 92152, USA*

**Abstract.** We refute a physical model, recently proposed by Gunn, Allison and Abbott (GAA) [http://arxiv.org/pdf/1402.2709v2.pdf], to utilize electromagnetic waves for eavesdropping on the Kirchhoff-law–Johnson-noise (KLJN) secure key distribution. Their model, and its theoretical underpinnings, is found to be fundamentally flawed because their assumption of electromagnetic waves violates not only the wave equation but also the Second Law of Thermodynamics, the Principle of Detailed Balance, Boltzmann's Energy Equipartition Theorem, and Planck's formula by implying infinitely strong blackbody radiation. We deduce the correct mathematical model of the GAA scheme, which is based on impedances at the quasi-static limit. Mathematical analysis and simulation results confirm our approach and prove that GAA's experimental interpretation is incorrect too.

## 1. Introduction

Recently Gunn, Allison and Abbott (GAA) [1] proposed a new, delay-based attack against the Kirchhoff-law–Johnson-noise (KLJN) secure key distribution scheme [2–5]. GAA claim—contradicting earlier statements most recently expounded in work by Kish, Abbott and Granqvist (KAG) [4]—that waves exist in a *finite-size* cable at *arbitrarily low frequencies*.

The theoretical basis of GAA's assertion [1] is the fact that, whereas wave-guides usually have a low-frequency cut-off for wave modes versus the *diameter* of the wave-guide, no such cut-off exists for transversal electromagnetic (TEM) wave modes in the case of *infinitely* long wave-guides. GAA write that, because coaxial cables have TEM wave modes, there is no frequency cut-off of the wave-based component of the electrical transport down to zero frequency. As a consequence of their presumption, GAA use the d'Alembert solution [1]

$$U(t,x) = U_{+}\left(t - \frac{x}{v}\right) + U_{-}\left(t - \frac{x}{v}\right) \tag{1}$$

for propagating lossless fluctuations—which may or may not be waves—in a linear medium to model the propagation of voltage in the cable used for key exchange in the KLJN scheme, where $U_{+}$ and $U_{-}$ are voltage components of waves propagating to the right and left along the $x$ axis, and $v$ is propagation velocity.

The experimental support of GAA's claim is that they have measured the voltage between the ends of a short coaxial cable at low-frequency sinusoidal voltage drive with an impedance-matched load at the other end and, at first sight, have found that their results confirmed some of the implications of Eq. 1, as further elaborated in Section 3 below.

We have analyzed GAA's statements [1] and found most of them invalid. Specifically, our findings and conclusions are the following:

(*i*) In cables, wave modes with wavelengths greater than twice their length are forbidden states, meaning that such modes do not exist; consequently there are no waves in cables in the frequency range pertinent to the KLJN scheme.

(*ii*) Instead, time-dependent propagation processes are non-wave type *retarded potentials* in a distributed impedance system; one of the implications of this is that Eq. 1 does not hold.

(*iii*) GAA's interpretation of their own "wave-verification" experiments [1] is invalid.

(*iv*) A correct derivation of the problem posed by GAA allows a proof that their attack does not work and an identification of the source of errors in their "attack-demonstration" experiments.

(*v*) We prove that directional coupler attacks under the KLJN conditions can never offer more information leak than the old wire resistance attacks. It should be noted that directional couplers in KLJN can never separate Alice's and Bob's contribution.

(*vi*) Finally we are able to formulate a general proof that any future (not yet existing) advanced attack type, that may be able even to *separate* the contribution of Alice and Bob, can never crack the KLJN scheme. Instead such an attack would lead to more advanced KLJN schemes that would not only be faster than earlier versions but would also be immune against these types of attacks.

The results delineated in (*i*) – (*vi*) will be derived and presented in three separate publications. The present paper is focused on items (*i*) – (*iii*), and moreover points out that the erroneous wave picture at low frequencies, referred to above, seems to be a common misconception in electrical engineering education.

**2. Refutation of GAA's theory and experimental interpretation**

As mentioned above, GAA's attack [1] on the KLJN scheme employs waves and related delays in a cable to extract information. While attempts to utilize time delays in cables for information purposes are to be encouraged, the asserted use of waves, which do not exist, is a fundamental flaw that invalidates GAA's basic considerations, proposed experiments, and interpretation of these experiments.

The wavelengths corresponding to the frequency range of concern in the KLJN scheme are much longer than the physical extent of the cable, and we earlier referred to that range the "no-wave" or "quasi-static" limit [2–5]. As remarked above, GAA argue that TEM wave modes do not exhibit any low-frequency cut-off. It is true that TEM wave modes in a wave-guide do not have a low-frequency cut-off *versus the diameter* of the wave-guide, but this argument is irrelevant because wave modes do have a cut-off *versus the length* of the cable. This does not imply that the electrical transport itself has a cut-off; it solely means that, when wave modes are forbidden, electrical transport takes place via non-wave phenomena—such as drift and relaxation—which constitute the form of transport in the quasi-static region of electrodynamics.

In this section, we use three separate fields of physics and mathematics to prove, in six different ways, that GAA's assumption of the existence of waves in the short cable within the frequency range pertinent to the KLJN scheme violates the following equations and laws of physics (our proofs are given subsequently):

(a) the wave equation,

(b) essential dynamical behavior of magnetic and electric energy in electromagnetic wave modes,

(c) Boltzmann's Energy Equipartition Theorem of Thermodynamics,

(d) the Principle of Detailed Balance of Thermodynamics,

(e) the Second Law of Thermodynamics, and

(f) theory and experiments for blackbody radiation.

*2.1 Violation of the wave equation*

It was recently pointed out by KAG [4] (including one of the proponents of the GAA model) that the wave equation precludes the existence of waves in the frequency range of concern for the KLNJ scheme. Next we provide more details about this fact and first illustrate the distributed inductance–capacitance–resistance (LCR) model of the cable in the KLJN scheme in Figure 1.

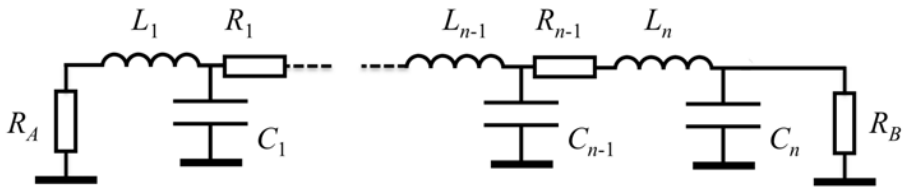

**Fig. 1.** Outline of the pertinent part of the KLJN scheme with a distributed LCR model of a long and leakage-free cable. When the cable losses can be neglected, one may omit the $R_i$ resistors representing the distributed resistance of the cable. Alice's and Bob's resistors, denoted $R_A$ and $R_B$, respectively, are randomly selected from the set $\{R_L, R_H\}$ with $(R_L \neq R_H)$ at the beginning of each bit-exchange period. These resistors, with associated serial generators (not shown), emulate thermal noise with high noise temperature and strongly limited bandwidth.

For the sake of simplicity but without losing generality, we discuss the case of a lossless cable. The main conclusion about the lack of wave modes is general because the inclusion of damping terms in harmonic differential equations *can never produce new eigen-frequencies*; they can only modify them and their bandwidth.

The wave equations of voltage $U(x,t)$ and current $I(x,t)$ in lossless cables are

$$\frac{\partial^2 U(x,t)}{\partial x^2} = \frac{1}{v_c^2} \frac{\partial^2 U(x,t)}{\partial t^2} \quad , \tag{2}$$

$$\frac{\partial^2 I(x,t)}{\partial x^2} = \frac{1}{v_c^2} \frac{\partial^2 I(x,t)}{\partial t^2} \quad , \tag{3}$$

where the phase propagation velocity of waves in the cable is

$$v_c = \sqrt{\frac{1}{L_u C_u}} \quad . \tag{4}$$

Here $L_u$ and $C_u$ are inductance and capacitance "densities" of the cable (with units of *H/m* and *F/m*), *i.e.*, the unit-length (one-meter) cable inductance and capacitance.

The general classical-physical solutions of these equations in infinite ideal cables are superpositions of waves, with arbitrary frequency, propagating in positive and negative directions in accordance with the d'Alembert solution in Eq. 1. However, in a cable with finite length *D*, the frequency-space of solutions is quantized to discrete values so that integer multiples of the half-wavelength fit in the cable. Thus the wavelength $\lambda_{max}$ of the wave with the lowest frequency $f_{min}$ can be written as

$$\lambda_{max} = 2D \quad , \quad f_{min} = \frac{v_c}{2D} \quad . \tag{5}$$

Frequencies below $f_{min}$, down to zero frequency, constitute a forbidden band of wave states.

The KLJN key exchanger operation strictly requires for security that its frequency range satisfies the quasi-static condition, *i.e.*,

$$f \ll f_{min} \quad . \tag{6}$$

Thus the wave-based scheme and considerations of GAA for eavesdropping violate not only the wave equations in Eqs. 2 and 3, and its d'Alembert solution in Eq. 1, but also the related other fields of classical and quantum physics of waves, because such non-existent solutions are forbidden states.

We note, in passing, that if the wave-based treatment by GAA [1] were correct, we would not have quantization of atomic electron shells, a forbidden band (energy gap) would not exist in solid state physics, semiconductor devices would not work, and even chemistry would be non-existent or at least very different.

*2.2 Violation of the energy dynamics within wave modes*

It is easy to visualize the pertinent physical situation even without the wave equation. While the energy of the wave mode is constant, any electromagnetic wave signifies propagation and oscillatory "bouncing" of energy between magnetic and electric energy forms. To have a physical wave, both aspects are essential, and the propagation must take place via "bouncing" of the energy between its two forms. For example, an L–C resonator has its "bouncing" phenomenon between the electric and magnetic energy forms but lacks propagation, and hence this oscillation is not a wave.

Consider now a short cable satisfying Relation 6, which is driven by an ac voltage generator $U(t)$ at one end and closed with a resistance $R_w$ given by

$$R_w = \sqrt{\frac{L_u}{C_u}} \tag{7}$$

at the other end. Then the cable has equal amounts of energy in the electric and magnetic forms, as discussed below, and both of these energies are oscillating. However, these energies are not "bouncing" between the electric and magnetic forms but oscillate between these forms and the generator. This fact is obvious from the considerations that follow next.

Quasi-static conditions (Relation 6) imply that the time-dependent voltage and current are spatially homogeneous along the wire, *i.e.*,

$$U(x,t) \simeq U(t) , \tag{8}$$

$$I(x,t) \simeq I(t) . \tag{9}$$

Thus the electrical and magnetic energies, denoted $E_e(t)$ and $E_m(t)$, respectively, in the cable can be written

$$E_e(t) = \frac{1}{2} C_c U^2(t) , \tag{10}$$

$$E_m(t) = \frac{1}{2} L_c I^2(t) = \frac{1}{2} L_c \frac{U^2(t)}{R_w^2} , \tag{11}$$

where $C_c = DC_u$ and $L_c = DL_u$ are the capacitance and inductance of the whole cable, respectively, and the last term of Eq. 11 originates from the well-known fact that a cable of arbitrary length, closed with the wave impedance (resistance) $R_w$, has an input impedance of exactly $R_w$. Thus

$$\frac{E_e(t)}{E_m(t)} = \frac{C_c}{L_c} R_w^2 = \frac{C_u}{L_u} R_w^2 = 1 \tag{12}$$

at each instant of time. Obviously, no "bouncing" phenomenon takes place between the electric and magnetic forms of energy because the energy is bouncing between the generator and the magnetic and electric fields, not between each other, as would be expected for a wave.

Now suppose that $R_A$ and $R_B$, rather than $R_w$, terminate the cable ends. The only significant change that will occur in the quasi-static limit is that the ratio of electrical and magnetic energies will be different from unity, which further goes against the wave hypothesis.

In conclusion, even the simple picture of wave modes given above proves that waves do not exist in the cable in the quasi-static limit.

*2.3 Violation of thermodynamics*

Simple but fundamental thermodynamic considerations also prove that waves cannot exist in the cable in the quasi-static frequency range.

*2.3.1 Violation of the Energy Equipartition Theorem, the principle of Detailed Balance and the Second Law of Thermodynamics*

According to Boltzmann's *Energy Equipartition Theorem* [6] for thermal equilibrium at temperature *T*, each electromagnetic wave mode has a mean thermal energy equal to *kT*, where *k* is Boltzmann's constant. Half of this mean energy is electrical and the other half is magnetic. For *N* different wave modes in the system, both the electrical and magnetic fields carry a total energy equal to *NkT*/2.

It is easy to see that the assumption that, in a *hypothetical* wave system, these wave energies are less than the above given values violates not only the *Energy Equipartition Theorem* but also the *Principle of Detailed Balance* [6] and the *Second Law of Thermodynamics*: Coupling this hypothetical wave system to a regular one would hit the *Detailed Energy Balance* of equilibrium between wave modes in the hypothetical and regular systems because it would yield an energy flow toward the hypothetical one. This energy flow could then be utilized for perpetual motion machines of the second kind, *i.e.*, violate the *Second Law of Thermodynamics*.

*2.3.2. Violation of Planck's Law and experimental facts for blackbody radiation*

Planck [7] deduced his law of thermal radiation, for simplicity, from the properties of a box with black walls, *i.e.*, internal walls with unity absorptivity and emissivity. In thermal equilibrium, thermal radiation within an infinitely large box with walls of arbitrary absorptivity, emissivity and color has a power spectral intensity for each polarization given by

$$I(f) = \frac{4\pi h f^3}{c^2} \frac{1}{e^{hf/kT} - 1} \quad , \tag{13}$$

where *h* is Planck's constant. The derivation of this formula is based on counting existing wave modes in the frequency range $f > f_{min}$, where the minimum frequency is obtained by the same frequency quantization as we claim to exist in a cable. GAA's notions (as well as common electrical engineering education) imply that in closing the finite-size cable by the wave impedance (resistance) $R_w$ at its two ends, all of the lower frequencies, at $f < f_{min}$, will also be available for wave modes as a consequence of the unity absorptivity and emissivity of the impedance match. However, one must realize that a cable closed by the wave impedance (resistance) $R_w$ at its two ends is a one-dimensional realization of Planck's box with perfectly absorbing (black) walls. Allowing the $f < f_{min}$ frequency range for wave modes results in a non-quantized continuum distribution of wave modes, which yields an infinite number *N* of wave modes in any finite frequency band. Such a situation in the finite-size cable results in an infinite amount of thermal energy *NkT* in any finite frequency band within the classical-physical frequency range $f \ll kT/h$. This situation is not a problem in a cable or box with infinite size. However, in a finite-size box, infinite thermal energy in the sum of wave modes would naturally result in infinite intensity of blackbody radiation.

Furthermore, the situation is similar for the KLJN scheme in which the cable ends are terminated by $R_A$ and $R_B$ rather than $R_w$. According to Planck's results, discussed above, the thermal radiation field in the closed box and cable does not depend on the absorptivity and emissivity of the walls—*i.e.*, on the cable termination resistances—since otherwise Planck's Law would be invalid and his formula violate the Second Law of Thermodynamics. Therefore the above argumentation regarding infinite energies and infinite thermal radiation holds for arbitrary termination and wall color.

If there is a loss in the cable and the cable is terminated by $R_A$ and $R_B$, the same situation holds provided $R_A$ and $R_B$, and the cable, have the same (noise) temperature because the system is still in thermal equilibrium.

In conclusion, GAA's claims violate the *Planck formula* and *experimental facts about blackbody radiation*.

*2.3.3 Proof that the total energy of the cable is less than the energy required for a single wave mode*

To further support the above conclusions we show that in the quasi-static frequency limit, wherein the KLJN scheme operates, the total energy is not sufficient even to supply a single wave mode, and consequently waves cannot exist without violating several principles and one law of thermodynamics.

As an example, we mathematically analyze how the no-wave situation manifests itself in a lossless cable, which is closed at both ends by resistors equal to the wave impedance $R_w$ at the temperature *T*. These conditions are not necessary but serve to simplify the calculations. Suppose that, in accordance with the KLJN conditions, the Johnson noise of the resistors has an upper cut-off frequency at $f_c$ so that

$$f_c \ll f_{min} = \frac{v_c}{2D} \tag{14}$$

hold in order to satisfy Relation 6. Thus the spatial homogeneity of current and voltage (Eqs. 8 and 9) is valid for the wire, and it is straightforward to calculate the electric and magnetic energies of thermal origin as shown next.

According to the Johnson–Nyquist formula [8], the thermal electrical energy in the cable capacitance is

$$E_{e,th} = \frac{1}{2}C_c \langle U^2(t) \rangle = \frac{1}{2}C_c \int_0^{f_c} \frac{4kTR_w/2}{1+f^2/f_{0C}^2}df \cong$$
$$\cong kTC_c R_w f_c = \frac{kT}{2}\frac{f_c}{f_{min}} \ll \frac{kT}{2} \tag{15}$$

and the thermal magnetic energy in the cable inductance is

$$E_{m,th} = \frac{1}{2}L_c \langle I^2(t) \rangle = \frac{1}{2}L_c \int_0^{f_c} \frac{4kT/(2R_w)}{1+f^2/f_{0L}^2}df \cong$$
$$\cong kT\frac{L_c}{R_w}f_c = \frac{kT}{2}\frac{f_c}{f_{min}} \ll \frac{kT}{2} \tag{16}$$

where the characteristic frequencies of the Lorentzian spectra in Eqs. 15 and 16 are defined as

$$f_{0C} = \frac{1}{2\pi C_c R_w/2} = \frac{1}{\pi D C_u}\sqrt{\frac{C_u}{L_u}} = \frac{1}{\pi D}\sqrt{\frac{1}{L_u C_u}} = \frac{1}{\pi}\frac{v_c}{D} = \frac{2}{\pi}f_{min} \tag{17}$$

and

$$f_{0L} = \frac{2R_w}{2\pi L_c} = \frac{1}{\pi L_c}\sqrt{\frac{L_u}{C_u}} = \frac{1}{\pi D}\sqrt{\frac{1}{L_u C_u}} = \frac{1}{\pi}\frac{v_c}{D} = \frac{2}{\pi}f_{min} \quad . \tag{18}$$

Similar calculations can be carried out for the general case in which the cable ends are not terminated by $R_w$ but with different resistance values $R_A$ and $R_B$. Specifically, the parallel resultant resistance $R_A$ and $R_B$ (instead of $R_w/2$) enters in the left-hand side of Eqs. 15 and 17, and the serial (sum) loop resistance $R_A+R_B$ (instead of $2R_w$) enters in the left-hand side of Eqs. 16 and 18, while the final inequalities shown by Eqs. 15 and 16 remain.

If there is a loss in the cable, and the cable is terminated by $R_A$ and $R_B$, the same situation holds provided $R_A$ and $R_B$, as well as the cable, have the same (noise) temperature, because the system is still in thermal equilibrium. If the cable is cooler then there is an energy flow out of the cable, which further strengthens the inequalities at the right hand sides of Eqs. 15 and 16.

In conclusion, Eqs. 15 and 16 prove that for a short cable and within the frequency range of interest for the KLJN scheme, the sum of electrical and magnetic energies in all of the hypothetical "wave modes" is much less than the energy needed for a single wave mode in thermal equilibrium.

**3. Correct treatment of cable delays in the frequency range for the KLJN scheme**

*3.1 General considerations*

We showed above that wave modes cannot exist in the cable at the KLJN condition $f << f_{min}$. A number of questions then arise naturally, such as (*i*) what type of system is the cable under these conditions, (*ii*) what is the nature of the propagating fluctuations caused by Alice's and Bob's noise generators, and (*iii*) are there any other implications of the KLJN condition?

To answer these questions, we first note that the system under consideration is not a waveguide, as implied by GAA [1], but a *distributed impedance network* in the *quasi-static limit*. Secondly, the propagating fluctuations are not waves but phase-shifted voltages and currents; in the language of physics they are related to *retarded potentials* of non-wave solutions, and in electrical engineering vocabulary they are *spatio-temporal fluctuations in an impedance network*.

The general implications of the KLJN conditions are very pervasive, as elaborated and discussed in Section 2 above. The specific consequences for the KLJN scheme are that the mathematical and physical framework used by GAA [1] is invalid and that the same applies to their experimental analysis, as shown in the present paper and as will be further discussed in forthcoming articles.

When the frequency converges towards zero, the impact of the inductance and capacitance of the cable on the cable current and voltage also rapidly diminish. However, the voltage drop over the cable is determined by its serial resistance $R_c$ and inductance $L_c$, because the capacitive shunt currents approach zero. Thus the first-order approximation of the cable impedance is

$$Z_c \cong R_c + j2\pi f L_c \ . \tag{19}$$

For simplicity, we analyze the situation wherein the cable loss (resistance) is negligible so that

$$Z_c \cong j2\pi f L_c . \tag{20}$$

The corresponding phase delay of the voltage at Bob's end, compared to that of Alice's end, is

$$\varphi_{AB} = -2\pi f L_c / R_B \tag{21}$$

when the voltage is generated by Alice. This phase delay corresponds to a *frequency-independent time delay* according to

$$\tau_{AB} = L_c / R_B \ , \tag{22}$$

which at first glance seems to suggest that we are dealing with waves and that the d'Alembert equation holds, as stated by GAA. However one must realize that this *time delay depends on the load resistance $R_B$* at the other end of the cable, which implies that the *time delay and measured phase velocity in the two directions are different* due to the condition $R_A \neq R_B$ during secure bit exchange [2–5], *i.e.*, under circumstances such that GAA's method [1] is supposed to function. To illustrate this dichotomy, we evaluate the phase delay for voltage propagation in the opposite direction, *i.e.*, when the voltage is generated by Bob. Now one finds

$$\varphi_{BA} = 2\pi f L_c / R_A \tag{23}$$

and

$$\tau_{BA} = L_c / R_A \ . \tag{24}$$

Thus the d'Alembert equation, applied by GAA [1] to prove the existence of waves, cannot be used in the present situation.

*3.2 Simulation based on a circuit model for the cable*

To corroborate the theoretical considerations above, we used Linear Technology's LTspice-IV cable simulator software to analyze the experimental situation in GAA's work [1] and confirmed all of their stated results. The simulations proved that a coaxial cable with parameters and conditions similar to those employed by GAA [1] can be modeled with the lumped impedance circuitry shown in Fig. 2, where parts (a) and (b) represent a cable with and without loss, respectively.

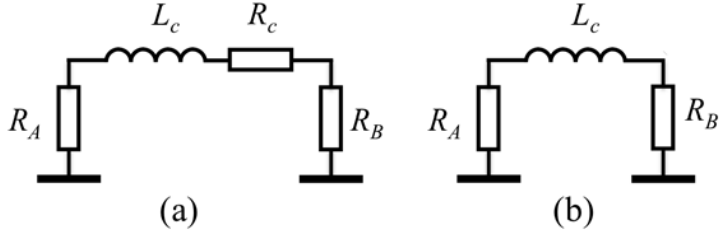

**Fig. 2.** Lumped impedance-components-based model of a cable at low frequencies for analyzing voltage drop along the cable and phase shift in the limit $f \ll f_{min}$. Part (a) represents a cable with loss (cable inductance and resistance are designated $L_c$ and $R_c$, respectively), and part (b) represents a lossless cable.

Figure 3 shows results of our simulation addressing the experimental data in GAA's article [1]. The conditions are the same as those of GAA [1] and reported in their Figure 5, but our simulation uses a practical cable model and simple impedance representations (see Figure 2) which fit the cable data to a high degree of accuracy. The practical lossy cable and the simple impedance model in Figure 2(a) give identical results, while results for the lossless cable (corresponding to data compensated for loss in GAA's work [1]) are nicely represented by the simple inductance model in Figure 2(b). Moreover, it is obvious that the cable inductance $L_c$ produces a voltage drop that is the time-derivative of the current, which is determined predominantly by the resistances in the loop. Thus the voltage drop for the lossless cable is the time derivative of Alice's generator voltage, and this experimental finding by GAA [1] to "support" the d'Alembert equation is simply an inductor-type voltage response and it has nothing to do with waves.

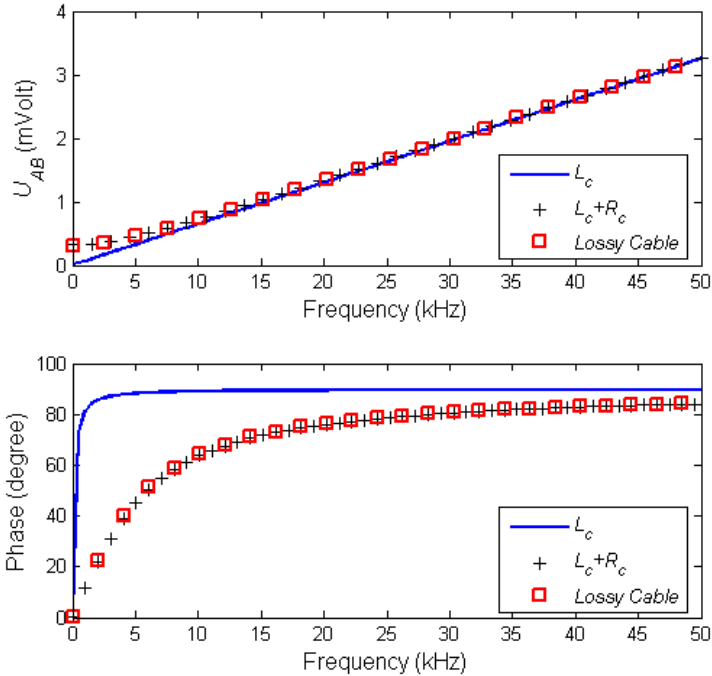

**Fig. 3.** Comparison of simulated data based on impedance models, using LTspice, with those of a real (lossy) cable. Alice's and Bob's resistors, denoted $R_A$ and $R_B$, have the resistance $R_W$, and Alice drives the cable with a sinusoidal voltage generator (1 V) via Bob's resistor. The cable is characterized by length $D = 1.5$ m, $L_c = 1.03$ µH, $R_c = 0.0315$ Ω, and $C_c = 150$ pF. Upper panel shows voltage drop $U_{AB}$ over the cable between Alice's and Bob's ends and lower panel shows phase shift of $U_{AB}$ compared to that of the voltage at Alice's end. Squares signify simulations of a lossy cable (model RG58), crosses represent data obtained by use of the lumped-parameters-model in Fig 2a, and solid line was derived from an inductance model devised to simulate a lossless

cable (Fig. 2b). These results are in full agreement with the experimental data shown in Figure 5 of GAA's work [1].

To subject Eqs. 22 and 24 to a final test, we evaluated the phase delay and corresponding time delay toward Bob when the resistor at Bob's end was varied and the cable was lossless. Data are shown in Table 1 and verify the correctness of our Eq. 22 to an accuracy of five digits. GAA's "propagation velocity" toward Bob is practically independent of frequency but depends on Bob's resistor. During secure bit exchange, the "propagation" times toward Alice and Bob are different. This fact verifies our conclusion that GAA's use of the d'Alembert equation, as the base of their mathematical considerations, is indeed incorrect.

**Table 1.** Simulated equivalent phase velocity calculated from phase shifts between the two ends of the cable versus driving frequency and load resistance (resistance of termination at the other end). The dependence on the resistance is in violation of the d'Alembert equation, for example, in the KLJN system during secure key exchange where the terminal resistances are different. Cable parameters are given in the caption for Fig 3.

|        | 1 kHz                       | 5 kHz                       |
|--------|-----------------------------|-----------------------------|
| **10 Ω**   | $3.99998 \times 10^7$ m/s   | $4.00018 \times 10^7$ m/s   |
| **20 Ω**   | $7.99996 \times 10^7$ m/s   | $8.00038 \times 10^7$ m/s   |
| **50 Ω**   | $1.99999 \times 10^8$ m/s   | $2.00007 \times 10^8$ m/s   |
| **1 kΩ**   | $3.99993 \times 10^9$ m/s   | $4.00011 \times 10^9$ m/s   |
| **10 kΩ**  | $3.99946 \times 10^{10}$ m/s | $4.00041 \times 10^{10}$ m/s |

One should observe that, for the cases of 1 kΩ and 10 kΩ, GAA's "phase velocity" is greater than the speed of light. This is acceptable and happens often with the phase velocity of oscillations in a *driven impedance system in the steady-state*; however, it is prohibited in the d'Alembert equation as a consequence of the theory of special relativity [9]. Similar effects can happen in wave-based systems with reflecting boundary conditions in stationary mode after the waves fill the system; however, in wave-based systems the phase velocity would be the same for the left and right directions.

In conclusion, the proper KLJN scheme is a simple impedance circuitry with related phase shifts where the corresponding time shifts are asymmetrical during secure bit exchange.

## 4. Additional notes

One of the reviewers raised an interesting question: In the light of the results of the present paper, how should one see the situation when a small part of a wave-based system is observed or when that part has different properties? Are there waves in those parts? Such a situation is common in physical systems and the answer is straightforward. We show two examples to elucidate the issue.

(a) First consider that a small part of a large wave-based system is observed, so small that Eq. 6 is satisfied. The answer to the question above is then simple: the part is a non-separable component of the wave system, and the observer views a small spatial fraction of the whole wave field. The energy of the observed wave mode is not the energy within this small part but the integral of related wave energy over the large system. However, if we separate the small system from the large one and terminate its ends, wave solutions would not exist at these frequencies.

(b) Another special example is a quarter-wavelength ($\lambda/4$) cable. If we terminate one end and drive the other end with a generator, the latter end acts as a distributed impedance transformer without wave modes. However if, instead of a generator, the same end is connected to a long cable with the same $R_w$, it will belong to the relevant wave mode that extends over the system and contains a small part of the energy of that wave mode.

## 5. Conclusions

The efforts of GAA [1] to utilize time delays in cables to crack the KLJN scheme represent an interesting and novel approach and, as such, deserve attention. However, it should be mentioned that Liu [10] earlier used a similar technique, but with unphysical conditions for the simulations [11]. As shown in considerable detail above, GAA's efforts can be irrevocably refuted. Further attempts to crack the KLJN scheme are welcome, but it is essential that they be founded on correct physical models.

Finally, we hope that the fact that waves cannot exist below $f_{min}$ in a short cable, as discussed in Section 2, will eventually enrich the physical underpinnings in electrical engineering courses and textbooks.

**Acknowledgements**

Discussions with Peter Hrasko, Robert Nevels, Lazar Kish, Krzysztof Michalski, Mehrdad Ehsani, Dave Ferry and Mark Dykman are appreciated. We are grateful to Lachlan Gunn for showing us the results in [1] and to Derek Abbott for discussions.